# An approach to non-homogenous Phase-type distributions through multiple cut-points

Juan Eloy Ruiz-Castro[1], Christian Acal[1], Juan B. Roldán[2]

[1]Department of Statistics and O.R. and IMAG, University of Granada, Spain

[2]Department of Electronics and Computer Technology, University of Granada, Spain

## Abstract

A new class of distributions based on phase-type distributions is introduced in the current paper to model lifetime data in the field of reliability analysis. This one is the natural extension of the distribution proposed by Acal et al. (2021) for more than one cut-point. Multiple interesting measures such as density function, hazard rate or moments, among others, were worked out both for the continuous and discrete case. Besides, a new EM-algorithm is provided to estimate the parameters by maximum likelihood. The results have been implemented computationally in R and simulation studies reveal that this new distribution reduces the number of parameters to be estimated in the optimization process and, in addition, it improves the fitting accuracy in comparison with the classical phase-type distributions, especially in heavy tailed distributions. An application is presented in the context of resistive memories with a new set of electron devices for non-volatile memory circuits. In particular, the voltage associated with the resistive switching processes that control the internal behaviour of resistive memories has been modelled with this new distribution to shed light on the physical mechanisms behind the operation of these memories.

**Keywords.** *Markov processes, phase-type distributions, multiple cut-points, modelling, EM-algorithm, simulation, Resistive Random Access Memory (RRAM)*

## 1. Introduction

In broad terms, reliability analysis aims to analyse the (random) lifetime of systems, being these subjected themselves to a continuous wear by several (un)controllable variables. Considering some probability distribution might shed light about the internal performance of systems and their main physical properties. In this context, the exponential distribution was the first distribution employed to ensure a reasonable fit for empirical data (Epstein and Sobel 1953). The exponential distribution has been the distribution par excellence during many years, but the assumption that units fail at constant rate (unrealistic situation) provoked the development and use of other probabilistic distributions such as Weibull, Gamma and Log-Normal, among others (Coolen 2008; McPherson 2013; Freels et al. 2019). While it is true that the behaviour of these distributions is acceptable for a wide range of real problems, the appearance of systems with more sophisticated internal structures spurs on to use other procedures that improve the quality of the fitting. Despite there are multiple options in the literature (see e.g. the combination of probability distributions in Kollu et al. 2012; Marques, Coelho, and de Carvalho 2015), a suitable candidate is to consider an approach based on Markov models. These models enable to modelling complex systems with well-structured results thanks to their algorithmic-matrix form. Besides, they make easier the posterior interpretation of results. Inside the theoretical framework of absorbent Markov processes, phase-type distributions (PH distributions) have an important role in the definition of new models (Titman and Sharples 2010; Titman 2014).

PH distributions were introduced by Neuts (1975, 1981) and verifies multiple good properties such as that this class is closed under a series of operations (maximum, minimum or addition) and generalizes a great number of well-known distributions. A detailed review about the main aspects of PH distributions can be seen in He (2014). Asmussen (2000) showed that the phase-type distribution class is dense in the non-negative probability distribution set. Therefore, any non-negative probability distribution can be estimated through a PH distribution with sufficient precision (Mahmoodi, Hamed-Ranjkesh, and Zhao 2020). Taking the power of PH distributions into account, they have been considered in many areas of knowledge such as queuing theory (Ramirez-Cobo, Lillo, and Wiper 2010), risk theory (Asmussen and Bladt 1996), medicine (Pérez-Ocón, Ruiz-Castro, and Gámiz-Pérez 1998; Ruiz-Castro and Zenga, 2020), reliability (Pérez-Ocón, Montoro-Cazorla, and Ruiz-Castro 2006) or electronics (Acal et al. 2019), among others.

However, PH distributions also show several disadvantages that complicate the optimization problem. On the one hand, PH distributions do not have a unique representation and on the other hand, the number of parameters to be estimated is usually high, especially in heavy tailed distributions or in distributions with multiple modes. Here, PH distributions do not always achieve a rigorous fit and then, it is easy to commit a misinterpretation of the reality. In order to solve these problems, it is crucial to develop new methodologies. Different proposals for heavy tailed distributions based on infinite-dimensional PH distributions and on certain transformations of PH distributions through inhomogeneous transitions rate can be seen in Bladt et al. (2014) and in Albrecher and Bladt (2019), respectively. Regarding reducing the number of parameters to be estimated, a solution is to consider a simpler structure where the dimension of estimation is smaller (Reinecke, Krauβ, and Wolter 2012). This idea would consist of assuming a well-known distribution (Erlang, Coxian, etc.) with their corresponding phase-type representation in the transition intensities matrix among transient states (Ausin, Wiper, and Lillo 2004; Gordon, Marshall, and Zenga 2018).

An alternative approach was proposed in Acal et al. (2021). In particular, a new distribution called one cut-point PH distribution was introduced to improve the quality of the fitting for similar scenarios. The underlying theoretical scheme is the following: let us suppose a data set with two different behaviours (see Figure 1). One cut-point, $a_1$, could be considered to determine the different zones. Then, the approach consists of considering two matrices, $\mathbf{T}_1$ and $\mathbf{T}_2$ with the same order and similar features than a matrix of a PH distribution, which describe the behaviour in each zone. This means that the internal behaviour of the states is not the same over time, i.e., the intensities are not the same in each interval. Therefore, one cut-point PH distribution can be seen as a first approach to non-homogeneous PH distributions that inherits the main characteristics of classical PH distributions. In this context, the current manuscript aims to generalize this distribution for the case with multiple cut-points.

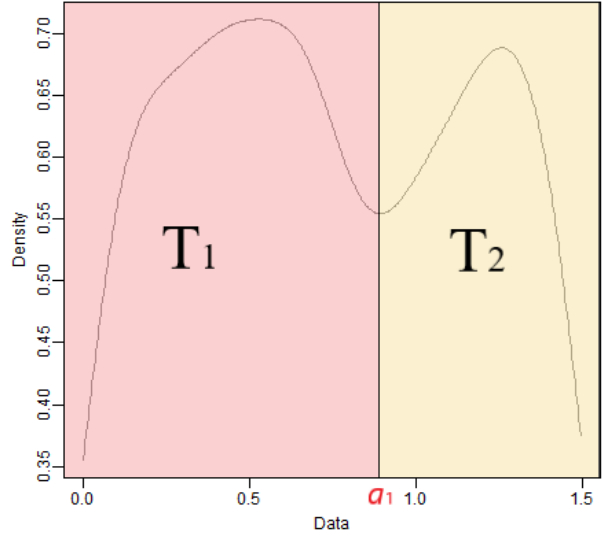

**Figure 1.** Empirical density function for a simulated data set. The theoretical scheme followed in the definition of one cut-point PH distribution is also displayed.

This new methodology can be applied to model the lifetime of any system with arbitrary accuracy. In particular, we focus on analysing data coming from resistive memories, also known as resistive random access memories (RRAMs). This kind of memories is one of the most promising technologies in the semiconductor industry, which is being introduced in different technological nodes in the industry to implement non-volatile memories. Recently, Lanza et al. (2022) reviewed the wide variety of applications where these devices will be key in the near future. Then, it is crucial to develop different mathematical approaches to describe and simulate the complex processes that occur in these devices (see, e.g., Mauri, Sacco, and Verri 2015; Ruiz-Castro et al. 2021), so that sector can understand better their operation and improve the quality of the modelling, which is essential for circuit design, where thousands of these devices are included in particular applications. In the current manuscript, the voltage in which RRAMs change their internal resistive state is analysed. In particular, most of these devices vary their conductance by creating and destroying a conductive filament that is formed within a dielectric and can allow the electrical connection of two metallic electrodes. The processes behind the creation and destruction of these conductive filaments are modelled through the multiple cut-points PH distribution (more information about the dynamics of the processes behind the conductive filaments can be checked in Ielmini and Waser 2015). The conductive filament evolution controls the conductivity of RRAMs, and therefore its behaviour within electronics circuits; hence, the importance of studying its behaviour is clear.

Besides this introduction, the rest of this document is organized as follows. In Section 2, a brief review about the one cut-point PH distribution is provided. The multiple cut-points PH distribution for the continuous and discrete cases are defined in Section 3 and Section 4, respectively. A specific EM algorithm to estimate the parameters of this new distribution together with a clarifying algorithm can be seen in Section 5. Besides, Section 6 contains several numerical examples where the potential of multiple cut-points PH distribution is displayed. A real application with RRAMs experimental data is shown in Section 7. Conclusions are given in Section 8.

## 2. A one continuous cut-point phase-type distribution

It is well-known that a continuous phase-type distribution can be defined as the time up to the absorption in an absorbing continuous Markov chain. Given a sub-conservative matrix $\mathbf{T}$ with order $m$, whose elements of the main diagonal are negative and the rest non-negative values where the sum of each row is less or equal to zero, and a vector $\boldsymbol{\alpha}$ with non-negative values whose sum is equal to one, then the variable $T$ follows a PH distribution with representation $(\boldsymbol{\alpha}, \mathbf{T})$ if its density function is expressed as

$$f(t) = \boldsymbol{\alpha} \exp(\mathbf{T}t) \mathbf{T}^0, \quad t \geq 0,$$

where the column vector $\mathbf{T}^0$ is defined as $\mathbf{T}^0 = -\mathbf{T}\mathbf{e}$ and $\mathbf{e}$ is a column vector of ones with appropriate order.

One approach to fit a distribution to a data set is to assume that the internal transitions behaviour among phases of the embedded Markov chain up to the absorption is different before and after a cut-point. That is, the Markov chain is non-homogeneous with one cut-point. Therefore, two matrices, $\mathbf{T}_1$ and $\mathbf{T}_2$, with the same order $m$ and similar features than a matrix of a PH distribution, which describe the behaviour in two areas in the non-negative real line, are considered.

**Definition.** A random variable $X$ is one cut-point phase-type distributed with representation $(\boldsymbol{\alpha}, \mathbf{T}_1, \mathbf{T}_2, a)$ defined in the non-negative real line if its density probability function is

$$f(x) = \begin{cases} \boldsymbol{\alpha} \exp(\mathbf{T}_1 x) \mathbf{T}_1^0 & ; \quad x \leq a \\ \boldsymbol{\alpha} \exp(\mathbf{T}_1 a) \exp(\mathbf{T}_2 (x-a)) \mathbf{T}_2^0 & ; \quad x > a, \end{cases}$$

for sub-conservative matrices $\mathbf{T}_1$ and $\mathbf{T}_2$ both with order $m$, whose elements of the main diagonal are negative and the rest non-negative values, and $\boldsymbol{\alpha}$ a row vector (order $m$) with non-negative values whose sum is equal to one.

From the density probability function, the cumulative and reliability probability functions can be worked out. The survival function has the following matrix structure

$$S(x) = \begin{cases} \boldsymbol{\alpha} \exp(\mathbf{T}_1 x) \mathbf{e} & ; \quad x \leq a \\ \boldsymbol{\alpha} \exp(\mathbf{T}_1 a) \exp(\mathbf{T}_2 (x-a)) \mathbf{e} & ; \quad x > a. \end{cases}$$

Details of this probability distribution can be seen in Acal et al. (2021). This probability distribution is generalized in this work.

## 3. Multiple cut-points phase-type distribution: the continuous case

The one cut-point case can be generalised by considering a general number $n$ of cut-points. The Markov chain associated with this distribution is non-homogeneous by considering the different areas partitioned by the cut-points.

**Definition.** Let $a_0 = 0 < a_1 < a_2 < ... < a_n < a_{n+1} = \infty$, $n$ cut-points ($a_0=0$ and $a_{n+1} = \infty$). Then, a random variable $X$ follows a continuous $n$-cut-points phase-type distribution with representation $(\boldsymbol{\alpha}, \mathbf{T}_1, ..., \mathbf{T}_{n+1}, a_1, ..., a_n)$ defined in the non-negative real line if the probability density function is given by

$$f(x) = \begin{cases} \boldsymbol{\alpha} \exp(\mathbf{T}_1 x) \mathbf{T}_1^0 & ; \quad a_0 = 0 < x \le a_1 \\ \boldsymbol{\alpha} \prod_{i=1}^{j-1} \exp(\mathbf{T}_i (a_i - a_{i-1})) \exp(\mathbf{T}_j (x - a_{j-1})) \mathbf{T}_j^0 & ; \quad a_{j-1} < x \le a_j \quad ; \quad 2 \le j \le n+1, \end{cases}$$

where all matrices $\mathbf{T}_i$ for $i = 1,..., n+1$ are sub-conservative matrices with identical order $m$. The elements of the main diagonal of these matrices are negative being positive or zero the remainder. Besides, the vector $\boldsymbol{\alpha}$ of order $m$ is a probability distribution and the matrix $\mathbf{T}_i^0$ is defined as $\mathbf{T}_i^0 = -\mathbf{T}_i \mathbf{e}$ for any $i$.

From the density probability function, several functions associated have been worked out, such as the cumulative probability function, $F(x) = P(X \le x)$, the reliability function $R(x) = P(X > x)$ and, from this, the cumulative hazard rate $H(x) = -\log R(x)$.

To calculate the moments of the distribution, the Laplace-transform

$$M_X(s) = E\left[e^{-sX}\right] = \int_0^\infty e^{-sx} f(x) dx,$$

is achieved. It is well known that the moments can be obtained from $M_X(s)$ by deriving this and evaluating in $s = 0$.

Appendix A shows these results in detail.

## 4. Multiple cut-points phase-type distribution: the discrete case

The motivation and definitions described in sections above can be developed for the discrete case. The discrete case is not an immediate consequence of the continuous case.

**Definition.** Let $a_0 = 0 < a_1 < a_2 < ... < a_n < a_{n+1} = \infty$, $n$ cut-points ($a_0=0$ and $a_{n+1} = \infty$). Then, a random variable $X$ follows a discrete $n$-cut-points phase-type distribution with representation $(\boldsymbol{\alpha}, \mathbf{T}_1, ..., \mathbf{T}_{n+1}, a_1, ..., a_n)$ if its probability mass function is

$$p_k = \begin{cases} \boldsymbol{\alpha} \mathbf{T}_1^{k-1} \mathbf{T}_1^0 & ;\ a_0 = 0 < k \leq a_1 \\ \boldsymbol{\alpha} \prod_{i=1}^{j-1} \mathbf{T}_i^{a_i - a_{i-1}} \mathbf{T}_j^{k - a_{j-1} - 1} \mathbf{T}_j^0 & ;\ a_{j-1} < k \leq a_j\ ;\ 2 \leq j \leq n+1. \end{cases}$$

In discrete case, all matrices $\mathbf{T}_i$ for $i = 1, ..., n+1$ are sub-stochastic matrices with identical order $m$. The elements of this matrix are non-negative; the vector $\boldsymbol{\alpha}$ of order $m$ is a probability distribution and the column vector $\mathbf{T}_i^0$ is defined as $\mathbf{T}_i^0 = (\mathbf{I} - \mathbf{T}_i)\mathbf{e}$ for any $i$.

Analogously to the continuous case, the cumulative probability function, $F_k = P(X \leq k) = \sum_{w \leq k} p_w$ and the reliability function are calculated.

To calculate the moments for a probability discrete distribution, the probability-generating function is built. This function is defined as

$$M_X(z) = E[z^X] = \sum_{k=1}^{\infty} p_k z^k,$$

where it converges absolutely at least for all complex numbers $z$ with $|z| \leq 1$; in many examples the radius of convergence is larger. From the derivation of this, the moments are obtained by evaluating in $z = 1$.

These functions and measures for the discrete case are given in detail in Appendix B.

## 5. Fitting a multiple cut-points phase-type distribution

One important problem is to fit a cut-points PH distribution to a data set. Let $(y_1, y_2, ..., y_q)$ be a finite independent observed sequence of data from a continuous cut-points PH distribution $Y$ with representation $(\boldsymbol{\alpha}, \mathbf{T}, \mathbf{a}) = (\boldsymbol{\alpha}, \mathbf{T}_1, ..., \mathbf{T}_{n+1}, a_1, ..., a_n)$ and an underlying non-homogeneous Markov chain $\{I(t); t \geq 0\}$ with $m$ transient states and one absorbing noted as $m+1$. The likelihood function according to the parameters of the distribution can be written as

$$L(\boldsymbol{\alpha}, \mathbf{T}, \mathbf{a} \mid y_1, ..., y_q) = \prod_{i=1}^{q} f(y_i \mid \boldsymbol{\alpha}, \mathbf{T}, \mathbf{a}).$$

The data set $(y_1, y_2, ..., y_q)$, includes the outcomes of $q$ independent replications of the cut-points PH distribution. For any $y_i$, the tuple $\mathbf{z}_i = (x_j^i, s_j^i; j = 0, ..., k_i - 1)$ describes the complete behaviour of the embedded non-homogeneous cut-point Markov process $\{I(t);$

$t \geq 0\}$ until absorption (states and sojourn time in each internal phase, respectively) being $k_i$ the number of states observed. Therefore, the observation $\mathbf{z} = \left(\mathbf{z}_i; i = 1,...,q\right)$ describes $q$ versions of visited phases and spent times till absorption. Then, the density of the complete sample $\mathbf{z}$ can be written as

$$f(\mathbf{z} \mid \boldsymbol{\alpha}, \mathbf{T}, \mathbf{a}) = \left(\prod_{i=1}^{m} \alpha_i^{B_i}\right)\left(\prod_{i=1}^{m}\prod_{h=1}^{n+1} \exp(Z_{i,h} t_{ii,h})\right)\left(\prod_{i=1}^{m}\prod_{\substack{j=1\\j\neq i}}^{m+1}\prod_{h=1}^{n+1} (t_{ij,h})^{N_{ij,h}}\right),$$

being $\alpha_i$ the initial probability of being initially in state $i$, $t_{ij,h}$ the transition intensity from $i$ to $j$ in the interval $(a_{h-1}, a_h]$ (element $(i, j)$ of the matrix $\mathbf{T}_h$), $B_i$ the number of Markov processes starting in state $i$, $Z_{i,h}$ the total time spent in state $i$ in the interval of time $(a_{h-1}, a_h]$ and $N_{ij,h}$ the total number of transitions jumps from $i$ to $j$ in the same interval of time $(a_{h-1}, a_h]$.

The density $f(\mathbf{z} \mid \cdot)$ is a member of a family with sufficient statistics

$$\mathbf{S} = \left((B_i)_{i=1,...,m}, (Z_{i,h})_{\substack{i=1,...,m\\h=1,...,n+1}}, (N_{ij,h})_{\substack{i=1,...,m\\j=1,...,m+1; i\neq j\\h=1,...,n+1}}\right),$$

and for this case it is well-known that the maximum likelihood estimates are (Basawa and Rao, 1980),

$$\hat{\alpha}_i = \frac{B_i}{q}; \quad \hat{t}_{ij,h} = \frac{N_{ij,h}}{Z_{i,h}}, \quad i \neq j; \quad \hat{t}_{im+1,h} = \frac{N_{im+1,h}}{Z_{i,h}}; \quad \hat{t}_{ii,h} = -\left(t_{im+1,h} + \sum_{j\neq i} t_{ij,h}\right).$$

The likelihood function is maximized by using an iterative method called EM-algorithm (Asmussen, Nerman, and Olsson 1996; Buchholz, Kriege, and Felko 2014) that works in two steps: expectation (E) and maximization (M). In particular, the following EM algorithm is built for estimating the parameters of multiple cut-points PH distributions.

## 5.1. The E- and the M- steps. Algorithming

In this development, it is assumed that cut-points are fixed beforehand. The first step of each iteration, the E-step, consist of calculating the conditional expectation, given the observed sample $\mathbf{y}$ and the current estimates of $(\boldsymbol{\alpha}, \mathbf{T}_1, ..., \mathbf{T}_{n+1})$, that is $\left(\boldsymbol{\alpha}^{(r)}, \mathbf{T}_1^{(r)}, ..., \mathbf{T}_{n+1}^{(r)}\right) \equiv (\boldsymbol{\alpha}, \mathbf{T}_1, ..., \mathbf{T}_{n+1})^{(r)}$.

The EM-algorithm to get the representation of the fitted distribution is given as follows.

*Step 0.* Input the seed distribution $\left(\boldsymbol{\alpha}^{(0)}, \mathbf{T}_1^{(0)}, ..., \mathbf{T}_{n+1}^{(0)}\right) \equiv \left(\boldsymbol{\alpha}, \mathbf{T}_1, ..., \mathbf{T}_{n+1}\right)^{(0)}$

*E-step.* Given the sample $\left(y_1, ..., y_q\right)$, let $B_i^{[v]}$, $Z_{i,h}^{[v]}$ and $N_{ij,h}^{[v]}$ be the contributions to $\mathbf{S}$ from the $v$ observation. The $r+1$ iteration is calculated as

$$B_i^{(r+1)} = \sum_{v=1}^{q} E_{(\boldsymbol{\alpha}, \mathbf{T}_1, ..., \mathbf{T}_{n+1})^{(r)}}\left(B_i^{[v]} \mid Y = y_v\right) = \sum_{v=1}^{q} \frac{\alpha_i^{(r)} \mathbf{e}_i^T \mathbf{b}\left(y_v \mid (\mathbf{T}_1, ..., \mathbf{T}_{n+1})^{(r)}\right)}{\boldsymbol{\alpha}^{(r)} \mathbf{b}\left(y_v \mid (\mathbf{T}_1, ..., \mathbf{T}_{n+1})^{(r)}\right)},$$

$$Z_{i,h}^{(r+1)} = \sum_{v=1}^{q} E_{(\boldsymbol{\alpha}, \mathbf{T}_1, ..., \mathbf{T}_{n+1})^{(r)}}\left(Z_{i,h}^{[v]} \mid Y = y_v\right) = \sum_{v=1}^{q} \frac{c_{ii}\left(y_v, h \mid (\boldsymbol{\alpha}, \mathbf{T}_1, ..., \mathbf{T}_h)^{(r)}\right)}{\boldsymbol{\alpha}^{(r)} \mathbf{b}\left(y_v \mid (\mathbf{T}_1, ..., \mathbf{T}_{n+1})^{(r)}\right)},$$

$$N_{ij,h}^{(r+1)} = \sum_{v=1}^{q} E_{(\boldsymbol{\alpha}, \mathbf{T}_1, ..., \mathbf{T}_{n+1})^{(r)}}\left(N_{ij,h}^{[v]} \mid Y = y_v\right) = \sum_{v=1}^{q} \frac{t_{ij,h}^{(r)} c_{ij}\left(y_v, h \mid (\boldsymbol{\alpha}, \mathbf{T}_1, ..., \mathbf{T}_h)^{(r)}\right)}{\boldsymbol{\alpha}^{(r)} \mathbf{b}\left(y_v \mid (\mathbf{T}_1, ..., \mathbf{T}_{n+1})^{(r)}\right)},$$

$$N_{im+1,h}^{(r+1)} = \sum_{v=1}^{q} E_{(\boldsymbol{\alpha}, \mathbf{T}_1, ..., \mathbf{T}_{n+1})^{(r)}}\left(N_{im+1,h}^{[v]} \mid Y = y_v\right)$$

$$= \sum_{v=1}^{q} \sum_{k=1}^{n+1} \frac{I_{\{a_{k-1} < y_v \leq a_k\}} t_{im+1,h}^{(r)} \mathbf{a}\left(y_v \mid (\boldsymbol{\alpha}, \mathbf{T}_1, ..., \mathbf{T}_{n+1})^{(r)}\right) \mathbf{e}_i}{\boldsymbol{\alpha}^{(r)} \mathbf{b}\left(y_v \mid (\mathbf{T}_1, ..., \mathbf{T}_{n+1})^{(r)}\right)}.$$

The vectors, $\mathbf{a}(y|\cdot)$, $\mathbf{b}(y|\cdot)$ and $c_{ij}(y,h|\cdot)$ contain; the probabilities of being in each transient state in the Markov process $I(\cdot)$ at time $y$ (row vector); the absorbing rates from the transient states at time $y$ (column vector); and the flow between transient states $i \to j$ in the interval $(a_{h-1}, a_h]$ and failure at time $y$, respectively. The vector $\mathbf{e}_i$ is a column vector of appropriate order whose elements are zero except the $i$-th which is equal to one and $\mathbf{e}_i^T$ denotes its transpose vector. The element $i$-th of the column vector $\mathbf{T}_h^0$ is denoted as $t_{im+1,h}$ (rate of failure from state $i$ in the interval $(a_{h-1}, a_h]$ in the embedded Markov process $I(\cdot)$).

The results of these conditioning expectations are demonstrated in Appendix C.

*M-step.* The values obtained in the E-step are used to generate a new set of parameters of the cut-points PH-distributions. The $r+1$ new estimates are given by

$$\alpha_i^{(r+1)} = \frac{B_i^{(r+1)}}{q}; \quad t_{ij,h}^{(r+1)} = \frac{N_{ij,h}^{(r+1)}}{Z_{i,h}^{(r+1)}}, \quad i \neq j; \quad t_{im+1,h}^{(r+1)} = \frac{N_{im+1,h}^{(r+1)}}{Z_{i,h}^{(r+1)}}; \quad t_{ii,h}^{(r+1)} = -\left(t_{im+1,h}^{(r+1)} + \sum_{j \neq i} t_{ij,h}^{(r+1)}\right)$$

**The algorithm**

**Input**: Data $(y_1,..., y_q)$

1. Choose the initial cut-point PH distribution $(\boldsymbol{\alpha}, \mathbf{T}_1,..., \mathbf{T}_{n+1})^{(0)}$
2. **Repeat**
    a. Compute for $r \geq 1$
        1. Compute for $y=(y_1,..., y_q)$ and $h = 1,…,n+1$
        $$\mathbf{a}\left(y\,|\,(\boldsymbol{\alpha}, \mathbf{T}_1,..., \mathbf{T}_{n+1})^{(r)}\right), \mathbf{b}\left(y\,|\,(\boldsymbol{\alpha}, \mathbf{T}_1,..., \mathbf{T}_{n+1})^{(r)}\right)$$
        $$c_{ij}\left(y, h\,|\,(\boldsymbol{\alpha}, \mathbf{T}_1,..., \mathbf{T}_{n+1})^{(r)}\right)$$
            a.1.1. *E-step*. Compute $B_i^{(r)}, Z_{i,h}^{(r)}, N_{ij,h}^{(r)}, N_{im+1,h}^{(r)}$
            a.1.2. *M-step*. Compute $\alpha_i^{(r)}$; $t_{ij,h}^{(r)}$; $t_{im+1,h}^{(r)}$; $t_{ii,h}^{(r)}$
    until $\left\|\boldsymbol{\alpha}^{(r)} - \boldsymbol{\alpha}^{(r-1)}\right\| + \sum_{h=1}^{n+1}\left\|\mathbf{T}_h^{(r)} - \mathbf{T}_h^{(r-1)}\right\| < \varepsilon$
3. **Output:** $\mathrm{PH}(\boldsymbol{\alpha}, \mathbf{T}_1,..., \mathbf{T}_{n+1})$

## 6. Numerical examples

In this section, several simulated datasets are modelled in order to show the potential and advantages of multiple cut-point phase-type distributions in comparison with the classical phase-type distributions. In practical studies, what is done is to consider different values of *n*, make a mesh of values for the cut-points and estimate the model parameters in each case using the EM algorithm described in the work.

### 6.1 Multi-modal distributions

A simulation problem has been analysed. A sample with size equal to 200 has been simulated of a multi-modal distribution by using the Gamma, Weibull and log-normal distribution. The empirical density function is given in Figure 2. In particular, the simulation process is the following:

1. Randomly, 200 sample points have been generated from Uniform(0,1) distribution, that is, $u_1, u_2, ..., u_{200}$.
2. For each $u_i$,
    i. If $u_i \leq 0.33$, a value from Gamma(20,0.1) distribution is generated.
    ii. If $0.33 < u_i \leq 0.66$, a value from Weibull(4,0.8) distribution is generated.
    iii. Otherwise, a value from Log-normal(1.2,0.08) distribution is generated.

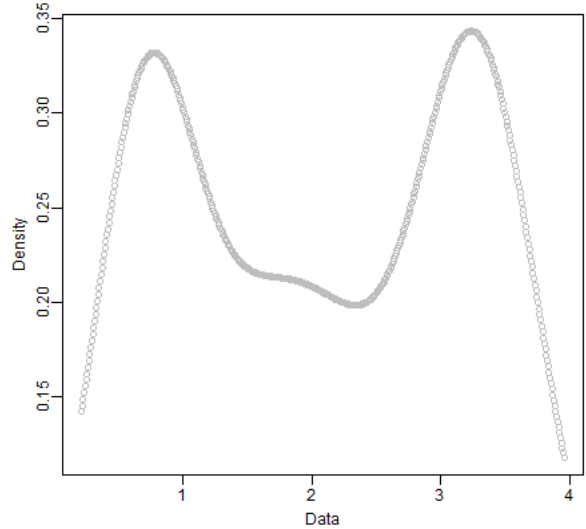

**Figure 2**. Empirical density distribution for the simulated dataset.

Firstly, different classical probability distributions have been employed without success to model this multi-modal distribution, not even Gamma, Weibull or Log-normal distribution (see Table 1). Other existing method is to apply some transformation on the data. However, this option should not be considered when the objective is to shed light about the behaviour and operation of systems. The transformation would be destroying the original structure of the data and would complicate the posterior interpretation of the parameters (Frost 2019).

**Table 1**. Summary for different classical probability distribution fits for the simulated dataset

| Distribution | Estimation | LogL | K-S test (p-value) | A-D test (p-value) |
|---|---|---|---|---|
| Normal | N(2.07,1.12) | -306.86 | 0.0027 | 0.0008 |
| Cauchy | C(2.09,0.92) | -373.88 | 0.0004 | 0.0001 |
| Logistic | LOG(2.07,0.70) | -318.19 | 0.0049 | 0.0010 |
| Exponential | EXP(0.48) | -345.38 | <0.0001 | <0.0001 |
| Gamma | Γ(2.65,1.28) | -304.26 | 0.0006 | 0.0005 |
| Log-normal | LN(0.53,0.69) | -315.25 | 0.0005 | 0.0002 |
| Weibull | W(1.91,2.33) | -298.25 | 0.0010 | 0.0006 |
| Gompertz | G(0.13,1.32) | -289.6 | <0.0001 | <0.0001 |

As it has been mentioned above, one of the main problems when multi-modal distributions are fitted through PH distributions is the number of necessary parameters for a rigorous adjust.

Initially, a PH distribution with 4 phases (before absorbing state) is fitted to the dataset by applying the classical EM algorithm developed by Asmussen et al. (1996). In total 24 parameters were estimated, being the logL value equal to −302.0914. The Kolmogorov-Smirnov (K-S) and Anderson-Darling (A-D) tests have been applied to check the goodness of fit. Since the obtained p-values were 0.00205 and 0.0022 for K-S and A-D tests, respectively, the fitted PH distribution must be rejected. The number of phases was incremented looking for getting the best fit but the number of parameters would increase considerably. To avoid this problem, several cut-points PH distributions introduced in this paper were adjusted.

The first approach has been to consider only one cut point and an Erlang structure (4 phases) in each interval of time. Then, the cut-point PH distribution considered has the following representation $(\boldsymbol{\alpha}, \mathbf{T}_1, \mathbf{T}_2, a_1)$ where

$$\boldsymbol{\alpha} = (1,0,0,0), \mathbf{T}_1 = \begin{pmatrix} -\lambda_1 & \lambda_1 & 0 & 0 \\ 0 & -\lambda_1 & \lambda_1 & 0 \\ 0 & 0 & -\lambda_1 & \lambda_1 \\ 0 & 0 & 0 & -\lambda_1 \end{pmatrix}, \mathbf{T}_2 = \begin{pmatrix} -\lambda_2 & \lambda_2 & 0 & 0 \\ 0 & -\lambda_2 & \lambda_2 & 0 \\ 0 & 0 & -\lambda_2 & \lambda_2 \\ 0 & 0 & 0 & -\lambda_2 \end{pmatrix}.$$

The EM-algorithm has been considered to estimate the model and the estimated parameters are $\lambda_1 = 2.8582, \lambda_2 = 1.4421$ and $a_1 = 0.82$. The maximum likelihood value is –291.4878 but the model must be rejected given that the p-values for K-S and A-D tests are 0.0008 and 0.0018, respectively.

The following step is to consider a two cut-points PH distribution with Erlang structure (also 4 phases) for the transition intensities matrices in each period of time. Here, the representation for this distribution is $(\boldsymbol{\alpha}, \mathbf{T}_1, \mathbf{T}_2, \mathbf{T}_3, a_1, a_2)$, where $\boldsymbol{\alpha}, \mathbf{T}_1$ and $\mathbf{T}_2$ adopt the same form than in the one cut-point distribution and

$$\mathbf{T}_3 = \begin{pmatrix} -\lambda_3 & \lambda_3 & 0 & 0 \\ 0 & -\lambda_3 & \lambda_3 & 0 \\ 0 & 0 & -\lambda_3 & \lambda_3 \\ 0 & 0 & 0 & -\lambda_3 \end{pmatrix}.$$

To estimate the parameters of this distribution, the EM-algorithm defined in Section 5 is assumed. After the estimation process, the method reaches the following optimum values: $\lambda_1 = 3.1609, \lambda_2 = 1.3633, \lambda_3 = 6.0570, a_1 = 0.49$ and $a_2 = 3.20$, with logL value equal to -270.7732. The adjust improves substantially, but it is not clear if this distribution should be accepted: p-value associated with K-S test is 0.0263, whereas A-D test provides a p-value equal to 0.0583.

We decide to increase the number of cut-points in three to guarantee a more accurate fit. Again, an Erlang structure with 4 phases is assumed in each interval of time and then, the three cut-points PH distribution has the following representation $(\boldsymbol{\alpha}, \mathbf{T}_1, \mathbf{T}_2, \mathbf{T}_3, \mathbf{T}_4, a_1, a_2, a_3)$, where $\boldsymbol{\alpha}, \mathbf{T}_1, \mathbf{T}_2, \mathbf{T}_3$ have the same structure than in the two cut-points distribution above and

$$\mathbf{T}_4 = \begin{pmatrix} -\lambda_4 & \lambda_4 & 0 & 0 \\ 0 & -\lambda_4 & \lambda_4 & 0 \\ 0 & 0 & -\lambda_4 & \lambda_4 \\ 0 & 0 & 0 & -\lambda_4 \end{pmatrix}.$$

After applying the EM-algorithm, the logL value is –251.7084, while the optimum values for the parameters of the distribution are $\lambda_1 = 2.6093, \lambda_2 = 2.8229, \lambda_3 = 1.0260, \lambda_4 = 5.6988$. Prior to the execution of EM-algorithm, the values of cut-points were fixed in $a_1 = 0.43, a_2 = 0.98$ and $a_3 = 3.15$ after optimizing the process under a given point grid. Taking the p-values related to K-S and A-D tests into account (0.6057 and 0.7026, respectively), this distribution can be accepted to model the experimental data considered in the current application.

Figure 3 shows the behaviour of the cumulative hazard rate for the different adjusted distributions. On the basis of the results, we can conclude that we improve significantly the fit through the new methodology and the number of parameters to estimate is reduced as well; 24 parameters in the classical PH approach and only 4 in the new one (PH distribution with three cut-points).

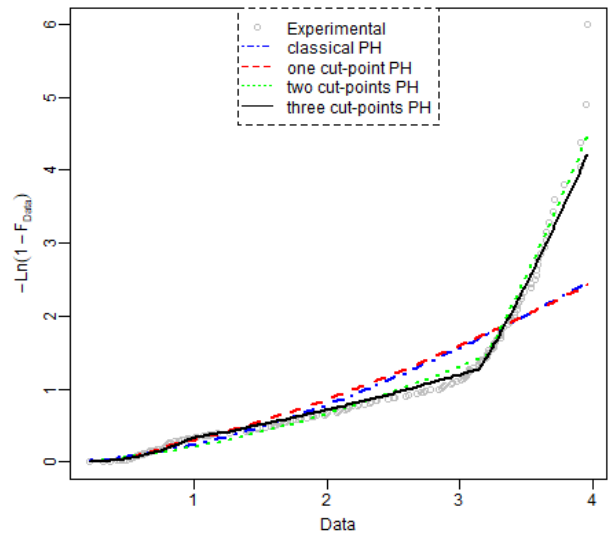

**Figure 3**. Experimental cumulative hazard rate for the simulated dataset and the corresponding fits.

### 6.2 Heavy-tailed distributions

A dataset with 200 sample points has been simulated from a Frechet distribution with values 0, 0.5 and 2 for location, scale and shape parameters, respectively. The empirical density function is given in Figure 4.

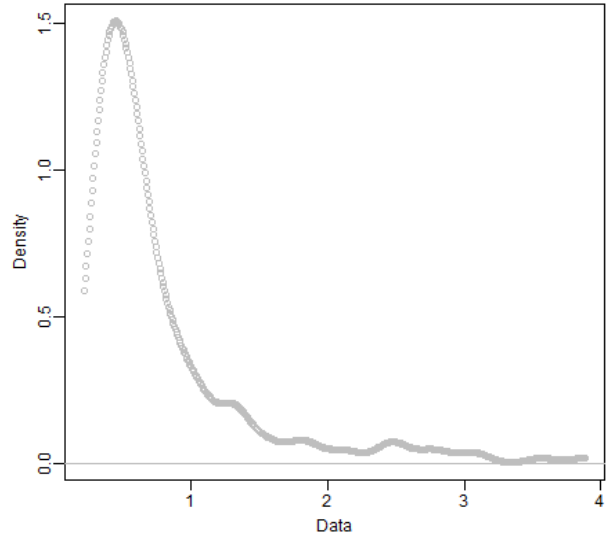

**Figure 4**. Empirical density distribution for the simulated dataset.

Initially, some classical probability distributions are considered to model this heavy tail distribution. According to the results obtained after applying the K-S and A-D goodness of fit tests, all of them must be rejected (see Table 2).

Table 2. Summary for different classical probability distribution fits for the simulated dataset

| Distribution | Estimation | LogL | K-S test (p-value) | A-D test (p-value) |
|---|---|---|---|---|
| Normal | N(0.79,0.64) | -195.35 | <0.0001 | <0.0001 |
| Cauchy | C(0.52,0.17) | -126.81 | <0.0001 | <0.0001 |
| Logistic | LOG(0.66,0.29) | -166.23 | <0.0001 | <0.0001 |
| Exponential | EXP(1.26) | -153.00 | <0.0001 | <0.0001 |
| Gamma | Γ(2.45,3.10) | -117.02 | 0.0003 | 0.0002 |
| Log-normal | LN(-0.45,0.61) | -94.5 | 0.0251 | 0.0175 |
| Weibull | W(1.44,0.88) | -131.12 | 0.0003 | <0.0001 |
| Gompertz | G(1.05,3.95) | -149.44 | <0.0001 | <0.0001 |

Then, a classical PH distribution is fitted by means of the EM-algorithm after assuming any internal structure for the transition intensities matrix. The optimum value is reached for 5 phases, while 35 parameters must be estimated for this distribution. This fit has been compared with the fit obtained by considering the multiple cut-points PH distributions in which an Erlang structure with 5 phases in each period of time has been assumed. In particular, one, two and three cut-points have been taken. Table 3 shows a brief summary about the estimation and goodness of the fit through K-S and A-D tests of each distribution. The experimental cumulative hazard rate estimated by these distributions have been plotted and compared graphically in Figure 5.

Table 3. Summary for the PH and multiple cut-point fits for the simulated dataset

| Distribution | Estimation | LogL | K-S test (p-value) | A-D test (p-value) |
|---|---|---|---|---|
| Classical PH (General structure with 5 phases) | ---------------- | −89.228 | 0.2273 | 0.0996 |
| One cut-point PH (Erlang structure with 5 phases) | $\lambda_1 = 8.0189$; $\lambda_2 = 2.4633$ $a_1 = 0.72$ | −84.556 | 0.8159 | 0.4348 |
| Two cut-point PH (Erlang structure with 5 phases) | $\lambda_1 = 8.0310$; $\lambda_2 = 3.0414$ $\lambda_3 = 1.8346$; $a_1 = 0.70$ $a_2 = 1.50$ | −82.641 | 0.8104 | 0.5082 |
| Three cut-point PH (Erlang structure with 5 phases) | $\lambda_1 = 8.0261$; $\lambda_2 = 3.2515$ $\lambda_3 = 1.5644$; $\lambda_4 = 2.8411$ $a_1 = 0.70$; $a_2 = 1.35$ $a_3 = 2.60$ | −80.893 | 0.8126 | 0.5403 |

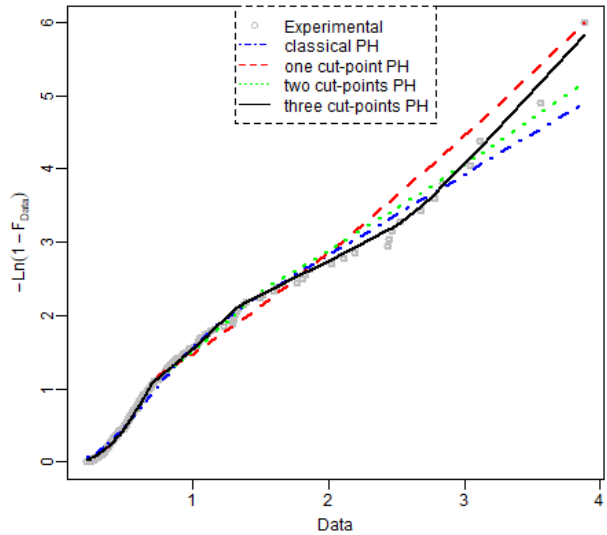

**Figure 5**. Experimental cumulative hazard rate for the simulated dataset and the corresponding fits

In relation to the analysis of the goodness of the fit, the classical PH distribution is accepted to model the simulated dataset. However, we can observe in Figure 5 (cumulative hazard rate) that this distribution reveals certain deficiencies in the distribution tail. This issue is solved by considering the cut-point PH distributions. As the number of cut-points increases, the accuracy of the fit improves significantly. In fact, with three cut-points all domain is practically under control. Therefore, the new class of distributions proposed in the current paper not only reduces the number of parameters to be estimated but also solves the lack of precision in the distribution tails.

# 7. Application

As stated above, RRAMs' operation is based on the resistive switching processes, which are characterized by the random creation and rupture of a conductive filament. These processes of formation and destruction are known as set and reset cycles, respectively. In particular, we aim to analyse the voltage at which this filament is formed ($V_{set}$) and broken ($V_{reset}$) in devices made of a dielectric made of $Al_2O_3$ (Cazorla et al. 2019). We have measured for one of these devices a resistive switching series of current versus voltages curves that include complete cycles where set and reset processes were considered. 1185 curves were obtained for both set and reset processes. From these curves we could extract the voltages at which the set ($V_{set}$) and reset ($V_{reset}$) events take place; these voltages constitute our datasets (the sample size equals 1185).

Firstly, Weibull distribution and classical PH distribution with Erlang structure were fitted to model $V_{set}$ and $V_{reset}$. The main reason why we make use of the Weibull distribution is because this distribution is usually considered in the context of RRAMs (Pan et al. 2014). However, neither of these distributions can be accepted according to Kolmogorov-Smirnov and Anderson-Darling tests (see Table 4).

Table 4. Summary for the Weibull and PH fits for the RRAMs datasets

| Dataset | Distribution | Estimation | LogL | K-S test (p-value) | A-D test (p-value) |
| --- | --- | --- | --- | --- | --- |
| $V_{set}$ | Weibull | $\beta = 4.536$ (shape) $\gamma = 0.1858$ (scale) | -1964.28 | <0.0001 | <0.0001 |
|  | Classical PH (Erlang structure with 15 phases) | $\lambda = 3.261$ | -1917.18 | 0.0271 | 0.0189 |
| $V_{reset}$ | Weibull | $\beta = 4.886$ (shape) $\gamma = 0.3525$ (scale) | −1116.46 | 0.0020 | 0.0045 |
|  | Classical PH (Erlang structure with 16 phases) | $\lambda = 6.150$ | −1158.60 | 0.0019 | 0.0003 |

Then, the multiple cut-point PH distribution is applied to these datasets in order to solve the lack of adjustment. Table 5 shows the goodness of the fit, whereas Figure 6 displays the experimental cumulative hazard rate and the corresponding fit for each dataset. We observe that the considered tests do not reject the distribution with a single cut-point, although we are not capable of controlling the adjustment in the tail distribution. Nevertheless, we can improve the quality of the fit in the tail distribution by considering the two cut-point PH distribution. Besides, the dimension of the transition intensities matrix is reduced again in each case.

Therefore, the fitting obtained with the new methodology introduced in this manuscript achieves the best results. Consequently, this finding allows to understand the structure of data and could help to unfold the physical mechanisms behind the resistive processes that control de operation of the devices. The potential application of this new distribution at the modelling context is clear, and could help to include the variability that takes place in a cycle to cycle manner in the design of new devices and the electronic circuits that can be built by including resistive memories.

Table 5. Summary for the multiple cut-point fits for the RRAMs datasets

| Dataset | Distribution | Estimation | LogL | K-S test (p-value) | A-D test (p-value) |
|---|---|---|---|---|---|
| $V_{set}$ | One cut-point PH (Erlang structure with 5 phases) | $\lambda_1 = 0.6151$<br>$\lambda_2 = 1.8470$<br>$a_1 = 3.33$ | -1897.93 | 0.3715 | 0.1910 |
| | Two cut-point PH (Erlang structure with 5 phases) | $\lambda_1 = 0.6191$<br>$\lambda_2 = 1.8258$<br>$\lambda_3 = 4.4817$<br>$a_1 = 3.33$<br>$a_2 = 7.65$ | -1889.05 | 0.3812 | 0.2892 |
| $V_{reset}$ | One cut-point PH (Erlang structure with 13 phases) | $\lambda_1 = 4.8787$<br>$\lambda_2 = 10.6958$<br>$a_1 = 3.20$ | -1095.25 | 0.4755 | 0.4816 |
| | Two cut-point PH (Erlang structure with 13 phases) | $\lambda_1 = 4.8826$<br>$\lambda_2 = 9.8197$<br>$\lambda_3 = 14.4195$<br>$a_1 = 3.20$<br>$a_2 = 3.55$ | -1092.80 | 0.5585 | 0.6422 |

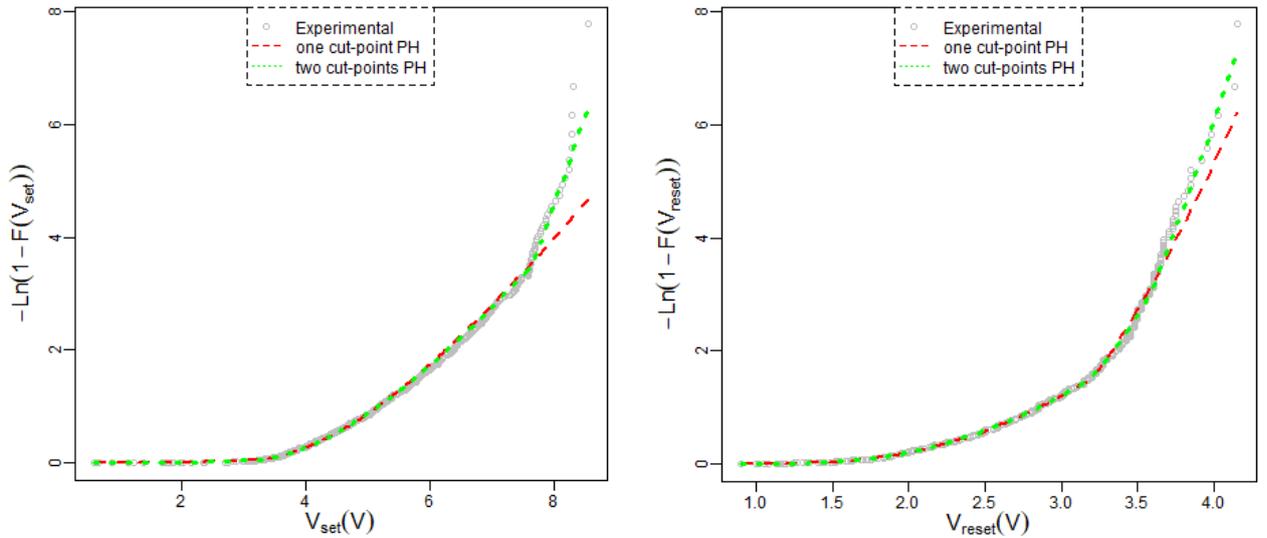

**Figure 6.** Experimental cumulative hazard rate for the RRAMs datasets under consideration, i.e., the set and reset voltage data, and the corresponding fits.

# 8. Conclusions

The study of reliability is an essential aspect in multiple areas of knowledge such as industry or medicine to understand better how systems work. Parametric probability distributions are normally used to guarantee an appropriate fit with empirical data, since they are fully known and enable to deduce solid conclusions on the basis of the interpretation of their parameters. However, classical probability distributions can reveal certain deficiencies when it comes to model experimental data from complex systems. On this matter, there exists a great need to propose new methodologies that ensure an accurate fitting. Many researchers are focusing their efforts on developing new approaches based on phase-type distributions. This class of distributions is dense in the set of probability distributions defined on any half-line of real numbers, so it is a reasonable aspirant to increase the quality of the fitting.

In the current manuscript a novel approach to non-homogenous phase-type distributions is introduced. In particular, a new class of distributions called multiple cut-points phase-type distributions is presented both for the continuous and discrete case of the associated Markov chain. This class of distributions can be seen as the natural extension of the methodology proposed in Acal et al. (2021) for more than one cut-point. Multiple cut-points phase-type distributions inherit the main properties of classical phase-type distributions. Besides, this class of distributions reduces the number of parameters to be estimated and improves the accuracy of the fit in comparison with other existing classical approaches, especially in those heavy tailed distributions or with multiple modes.

All methodology has been developed in an algorithmic-matrix way, which allows expressing and interpreting the results for any internal structure of the matrices involved. Then, the expressions worked out, as well as their interpretation, are the same for any number of cut-points.

A new algorithm (EM) has been built to estimate the parameters in a fit with cut-points phase-type distributions. This is an extension of what was given by Asmussen et al. (1996) for phase-type distributions. That is an iterative process to maximize a likelihood in the presence of missing data. From a theoretical point of view, the greater the number of cut-points, the better the fit, but at the same time the number of parameters will increase. In practice, if distributions such as Erlang's (one parameter per matrix) are considered, the fit obtained is already good enough with a minimum number of parameters. The applicability of this new method is shown in Sections 6 and 7.

All these characteristics are exposed through several simulated numerical examples in the paper. In addition, a real proof-of-concept of the appropriateness of these distributions is performed making use of experimental data obtained from resistive memories. The new modelling approach can be applied both to improve circuit design at the industrial level and to understand the physics behind the operation of resistive memories.

**Acknowledgements**

This paper is partially supported by the project FQM-307 of the Government of Andalusia (Spain) and by the project PID2020-113961GB-I00 of the Spanish Ministry of Science and Innovation (also supported by the FEDER programme). Authors also acknowledge the financial support of the Consejería de Conocimiento, Investigación y Universidad, Junta de Andalucía (Spain) and the FEDER programme for projects A-TIC-117-UGR18, A-FQM-66-UGR20 and B-TIC-624-UGR20. Additionally, the authors would like to express their gratitude for financial support by the IMAG–María de Maeztu grant CEX2020-001105-M/AEI/10.13039/501100011033.

**APPENDIX A**

The main results given in Section 3 for the continuous *n*-cut-points phase-type distribution are shown in this appendix. It is used in these proofs that $\mathbf{T}_i^0 = -\mathbf{T}_i \mathbf{e}$ and the existence of inverse matrices of $\mathbf{T}_i$ for $i = 1,\ldots, n+1$ given its structure.

*Cumulative Probability Function*

The cumulative probability function $F(x)$ from the density probability function is given by

$$F(x) = \begin{cases} 1 - \boldsymbol{\alpha} \exp(\mathbf{T}_1 x) \mathbf{e} & ;\ a_0 = 0 < x \leq a_1 \\ 1 - \boldsymbol{\alpha} \prod_{i=1}^{j-1} \exp(\mathbf{T}_i (a_i - a_{i-1})) \exp(\mathbf{T}_j (x - a_{j-1})) \mathbf{e} & ;\ a_{j-1} < x \leq a_j\ ;\ 2 \leq j \leq n+1 \end{cases}$$

Proof.

- Case $a_0 = 0 < x \leq a_1$

$$F(x) = \int_0^x \boldsymbol{\alpha} \exp(\mathbf{T}_1 u) \mathbf{T}_1^0 du = \boldsymbol{\alpha} \left(\exp(\mathbf{T}_1 x) - \mathbf{I}\right) \mathbf{T}_1^{-1} \mathbf{T}_1^0 = -\boldsymbol{\alpha} \left(\exp(\mathbf{T}_1 x) - \mathbf{I}\right) \mathbf{e} = 1 - \boldsymbol{\alpha} \exp(\mathbf{T}_1 x) \mathbf{e}.$$

- Case $a_{j-1} < x \leq a_j\ ;\ 2 \leq j \leq n+1$

$$F(x) = \sum_{k=1}^{j} B_k\text{, where}$$

$$B_1 = \int_0^{a_1} \boldsymbol{\alpha} \exp(\mathbf{T}_1 u) \mathbf{T}_1^0 du = 1 - \boldsymbol{\alpha} \exp(\mathbf{T}_1 a_1) \mathbf{e}$$

and for $k = 2, \ldots, j-1$ ($j \geq 3$)

$$B_k = \int_{a_{k-1}}^{a_k} \boldsymbol{\alpha} \prod_{i=1}^{k-1} \exp(\mathbf{T}_i (a_i - a_{i-1})) \exp(\mathbf{T}_k (u - a_{k-1})) \mathbf{T}_k^0 du$$

$$= \boldsymbol{\alpha} \prod_{i=1}^{k-1} \exp(\mathbf{T}_i (a_i - a_{i-1})) \mathbf{e} - \boldsymbol{\alpha} \prod_{i=1}^{k-1} \exp(\mathbf{T}_i (a_i - a_{i-1})) \exp(\mathbf{T}_k (a_k - a_{k-1})) \mathbf{e}$$

and

$$B_j = \int_{a_{j-1}}^{x} \boldsymbol{\alpha} \prod_{i=1}^{j-1} \exp(\mathbf{T}_i (a_i - a_{i-1})) \exp(\mathbf{T}_j (u - a_{j-1})) \mathbf{T}_j^0 du$$

$$= \boldsymbol{\alpha} \prod_{i=1}^{j-1} \exp(\mathbf{T}_i (a_i - a_{i-1})) \mathbf{e} - \boldsymbol{\alpha} \prod_{i=1}^{j-1} \exp(\mathbf{T}_i (a_i - a_{i-1})) \exp(\mathbf{T}_j (x - a_{j-1})) \mathbf{e}$$

Therefore,

$$F(x) = \sum_{k=1}^{j} B_k = 1 - \prod_{i=1}^{j-1} \exp(\mathbf{T}_i (a_i - a_{i-1})) \exp(\mathbf{T}_j (x - a_{j-1})) \mathbf{e}. \qquad \square$$

From the cumulative probability function, the reliability, $R(x) = P(X > x)$, and the cumulative hazard function, $H(x) = -\log R(x)$, are immediate.

*Laplace-Transform: $M_X(s)$*

The Laplace-transform function defined as $M_X(s) = E\left[e^{-sX}\right] = \int_0^\infty e^{-sx} f(x) dx$ for the $n$ cut-points phase-type distribution is given by

$$M_X(s) = -\boldsymbol{\alpha}(\mathbf{T}_1 - s\mathbf{I})^{-1} \mathbf{T}_1^0 + \boldsymbol{\alpha} \sum_{i=1}^{n} \left[ \prod_{j=0}^{i-1} e^{a_j(\mathbf{T}_j - \mathbf{T}_{j+1})} \right] e^{a_i(\mathbf{T}_i - s\mathbf{I})} \left( (\mathbf{T}_i - s\mathbf{I})^{-1} \mathbf{T}_i^0 - (\mathbf{T}_{i+1} - s\mathbf{I})^{-1} \mathbf{T}_{i+1}^0 \right)$$

for $s$ where the integral is convergent (and therefore the inverse of the matrices exits).

Proof.

Before the proof, it is noted that the probability density function defined in Section 3 can be expressed as follows

$$f(x) = \begin{cases} \boldsymbol{\alpha} \exp(\mathbf{T}_1 x) \mathbf{T}_1^0 & ; \; a_0 = 0 < x \leq a_1 \\ \boldsymbol{\alpha} \prod_{i=1}^{j-1} \exp(a_i (\mathbf{T}_i - \mathbf{T}_{i+1})) \exp(\mathbf{T}_j x) \mathbf{T}_j^0 & ; \; a_{j-1} < x \leq a_j \; ; \; 2 \leq j \leq n+1. \end{cases}$$

- Case $n = 1$

$$M_X(s) = E[e^{-sX}] = \int_0^\infty e^{-sx} f(x) dx$$
$$= \int_0^{a_1} \boldsymbol{\alpha} \exp(x(\mathbf{T}_1 - s\mathbf{I})) \mathbf{T}_1^0 dx + \int_{a_1}^\infty \boldsymbol{\alpha} \exp(a_1(\mathbf{T}_1 - \mathbf{T}_2)) \exp(x(\mathbf{T}_2 - s\mathbf{I})) \mathbf{T}_2^0 dx$$
$$= -\boldsymbol{\alpha}(\mathbf{T}_1 - s\mathbf{I})^{-1} \mathbf{T}_1^0 + \boldsymbol{\alpha} \exp(a_1(\mathbf{T}_1 - s\mathbf{I})) \left[ (\mathbf{T}_1 - s\mathbf{I})^{-1} \mathbf{T}_1^0 - (\mathbf{T}_2 - s\mathbf{I})^{-1} \mathbf{T}_2^0 \right]$$

- In general ($n \geq 2$) we have that, $M_X(s) = E[e^{-sX}] = \int_0^\infty e^{-sx} f(x) dx = \sum_{k=1}^{n+1} A_k$ where,

$$A_1 = \int_0^{a_1} e^{-sx} f(x) dx = \int_0^{a_1} \boldsymbol{\alpha} \exp(x(\mathbf{T}_1 - s\mathbf{I})) \mathbf{T}_1^0 dx$$
$$= -\boldsymbol{\alpha}(\mathbf{T}_1 - s\mathbf{I})^{-1} \mathbf{T}_1^0 + \boldsymbol{\alpha} \exp(a_1(\mathbf{T}_1 - s\mathbf{I}))(\mathbf{T}_1 - s\mathbf{I})^{-1} \mathbf{T}_1^0$$

$$A_2 = \int_{a_1}^{a_2} e^{-sx} f(x) dx = \int_{a_1}^{a_2} \boldsymbol{\alpha} \exp(a_1(\mathbf{T}_1 - \mathbf{T}_2)) \exp(x(\mathbf{T}_2 - s\mathbf{I})) \mathbf{T}_2^0 dx$$
$$= \boldsymbol{\alpha} \exp(a_1(\mathbf{T}_1 - \mathbf{T}_2)) \exp(a_2(\mathbf{T}_2 - s\mathbf{I}))(\mathbf{T}_2 - s\mathbf{I})^{-1} \mathbf{T}_2^0 - \boldsymbol{\alpha} \exp(a_1(\mathbf{T}_1 - s\mathbf{I}))(\mathbf{T}_2 - s\mathbf{I})^{-1} \mathbf{T}_2^0$$

For $i = 3, \ldots, n$ ($n \geq 3$)

$$A_k = \int_{a_{k-1}}^{a_k} e^{-sx} f(x) dx = \int_{a_{k-1}}^{a_k} \boldsymbol{\alpha} \prod_{j=1}^{k-1} \exp(a_j(\mathbf{T}_j - \mathbf{T}_{j+1})) \exp(x(\mathbf{T}_k - s\mathbf{I})) \mathbf{T}_k^0 dx$$
$$= \boldsymbol{\alpha} \prod_{j=1}^{k-1} \exp(a_j(\mathbf{T}_j - \mathbf{T}_{j+1})) \exp(a_k(\mathbf{T}_k - s\mathbf{I}))(\mathbf{T}_k - s\mathbf{I})^{-1} \mathbf{T}_k^0$$
$$- \boldsymbol{\alpha} \prod_{j=1}^{k-2} \exp(a_j(\mathbf{T}_j - \mathbf{T}_{j+1})) \exp(a_{k-1}(\mathbf{T}_{k-1} - s\mathbf{I}))(\mathbf{T}_k - s\mathbf{I})^{-1} \mathbf{T}_i^0$$

$$A_{n+1} = \int_{a_n}^\infty e^{-sx} f(x) dx = \int_{a_n}^\infty \boldsymbol{\alpha} \prod_{k=1}^n \exp(a_j(\mathbf{T}_j - \mathbf{T}_{j+1})) \exp(x(\mathbf{T}_{n+1} - s\mathbf{I})) \mathbf{T}_{n+1}^0 dx$$
$$= -\boldsymbol{\alpha} \prod_{j=1}^{n-1} \exp(a_j(\mathbf{T}_j - \mathbf{T}_{j+1})) \exp(a_n(\mathbf{T}_n - s\mathbf{I}))(\mathbf{T}_{n+1} - s\mathbf{I})^{-1} \mathbf{T}_{n+1}^0$$

From these expressions we can observe that,

$$\sum_{i=1}^n A_n = -\boldsymbol{\alpha}(\mathbf{T}_1 - s\mathbf{I})^{-1} \mathbf{T}_1^0 + \boldsymbol{\alpha} \sum_{i=1}^{n-1} \left[ \prod_{j=0}^{i-1} e^{a_j(\mathbf{T}_j - \mathbf{T}_{j+1})} \right] e^{a_i(\mathbf{T}_i - s\mathbf{I})} \left( (\mathbf{T}_i - s\mathbf{I})^{-1} \mathbf{T}_i^0 - (\mathbf{T}_{i+1} - s\mathbf{I})^{-1} \mathbf{T}_{i+1}^0 \right)$$
$$+ \boldsymbol{\alpha} \prod_{j=0}^{n-1} e^{a_j(\mathbf{T}_j - \mathbf{T}_{j+1})} \exp(a_n(\mathbf{T}_n - s\mathbf{I}))(\mathbf{T}_n - s\mathbf{I})^{-1} \mathbf{T}_n^0.$$

Finally,

$$M_X(s) = \sum_{k=1}^{n+1} A_k$$

$$= -\boldsymbol{\alpha}(\mathbf{T}_1 - s\mathbf{I})^{-1}\mathbf{T}_1^0 + \boldsymbol{\alpha}\sum_{i=1}^{n}\left[\prod_{j=0}^{i-1} e^{a_j(\mathbf{T}_j - \mathbf{T}_{j+1})}\right] e^{a_i(\mathbf{T}_i - s\mathbf{I})}\left((\mathbf{T}_i - s\mathbf{I})^{-1}\mathbf{T}_i^0 - (\mathbf{T}_{i+1} - s\mathbf{I})^{-1}\mathbf{T}_{i+1}^0\right)$$

□

*Expectation: E[x]*

It is well-known that $E[X^m] = (-1)^m \left.\dfrac{\partial^m M_X(s)}{\partial s^m}\right|_{s=0}$. Therefore, $E[X] = -\left.\dfrac{\partial M_X(s)}{\partial s}\right|_{s=0}$.

$$\frac{\partial M_X(s)}{\partial s} = -\boldsymbol{\alpha}(\mathbf{T}_1 - s\mathbf{I})^{-2}\mathbf{T}_1^0 - \boldsymbol{\alpha}\sum_{i=1}^{n}\left[\prod_{j=0}^{i-1} e^{a_j(\mathbf{T}_j - \mathbf{T}_{j+1})}\right] a_i e^{a_i(\mathbf{T}_i - s\mathbf{I})}\left((\mathbf{T}_i - s\mathbf{I})^{-1}\mathbf{T}_i^0 - (\mathbf{T}_{i+1} - s\mathbf{I})^{-1}\mathbf{T}_{i+1}^0\right) +$$

$$+ \boldsymbol{\alpha}\sum_{i=1}^{n}\left[\prod_{j=0}^{i-1} e^{a_j(\mathbf{T}_j - \mathbf{T}_{j+1})}\right] e^{a_i(\mathbf{T}_i - s\mathbf{I})}\left((\mathbf{T}_i - s\mathbf{I})^{-2}\mathbf{T}_i^0 - (\mathbf{T}_{i+1} - s\mathbf{I})^{-2}\mathbf{T}_{i+1}^0\right)$$

$$E[X] = -\left.\frac{\partial M_X(s)}{\partial s}\right|_{s=0} = \boldsymbol{\alpha}\mathbf{T}_1^{-2}\mathbf{T}_1^0 + \boldsymbol{\alpha}\sum_{i=1}^{n}\left[\prod_{j=0}^{i-1} e^{a_j(\mathbf{T}_j - \mathbf{T}_{j+1})}\right] a_i e^{a_i\mathbf{T}_i}\left(\mathbf{T}_i^{-1}\mathbf{T}_i^0 - \mathbf{T}_{i+1}^{-1}\mathbf{T}_{i+1}^0\right)$$

$$- \boldsymbol{\alpha}\sum_{i=1}^{n}\left[\prod_{j=0}^{i-1} e^{a_j(\mathbf{T}_j - \mathbf{T}_{j+1})}\right] e^{a_i\mathbf{T}_i}\left(\mathbf{T}_i^{-2}\mathbf{T}_i^0 - \mathbf{T}_{i+1}^{-2}\mathbf{T}_{i+1}^0\right)$$

$$= -\boldsymbol{\alpha}\mathbf{T}_1^{-1}\mathbf{e} + \boldsymbol{\alpha}\sum_{i=1}^{n}\left[\prod_{j=0}^{i-1} e^{a_j(\mathbf{T}_j - \mathbf{T}_{j+1})}\right] a_i e^{a_i\mathbf{T}_i}\left(-\mathbf{e} + \mathbf{e}\right)$$

$$- \boldsymbol{\alpha}\sum_{i=1}^{n}\left[\prod_{j=0}^{i-1} e^{a_j(\mathbf{T}_j - \mathbf{T}_{j+1})}\right] e^{a_i\mathbf{T}_i}\left(-\mathbf{T}_i^{-1}\mathbf{e} + \mathbf{T}_{i+1}^{-1}\mathbf{e}\right)$$

$$= -\boldsymbol{\alpha}\mathbf{T}_1^{-1}\mathbf{e} + \boldsymbol{\alpha}\sum_{i=1}^{n}\left[\prod_{j=0}^{i-1} e^{a_j(\mathbf{T}_j - \mathbf{T}_{j+1})}\right] e^{a_i\mathbf{T}_i}\left(\mathbf{T}_i^{-1} - \mathbf{T}_{i+1}^{-1}\right)\mathbf{e}.$$

□

Similarly, the second order moment can be calculated as,

$$E\left[X^2\right] = 2\boldsymbol{\alpha}\mathbf{T}_1^{-2}\mathbf{e} - 2\boldsymbol{\alpha}\sum_{i=1}^{n}\left[\prod_{j=0}^{i-1}e^{a_j(\mathbf{T}_j - \mathbf{T}_{j+1})}\right]e^{a_i\mathbf{T}_i}\left[\mathbf{T}_{i+1}^{-1}\left(a_i\mathbf{I} - \mathbf{T}_{i+1}^{-1}\right) - \mathbf{T}_i^{-1}\left(a_i\mathbf{I} - \mathbf{T}_i^{-1}\right)\right]\mathbf{e}.$$

**APPENDIX B**

The discrete case is given in Section 4. The main results for the discrete *n*-cut-points phase-type distribution are shown in this appendix. We use in these proofs that in the discrete case $\mathbf{T}_j^0 = (\mathbf{I} - \mathbf{T}_j)\mathbf{e}$ and the matrix $\mathbf{I} - \mathbf{T}_j$ is non-singular for $j = 1,\ldots, n+1$ given the internal structure of these matrices.

*Cumulative Probability Function: $F_k$*

From the probability mass function given in Section 4 the cumulative distribution function is

$$F_k = P(X \leq k) = \sum_{w \leq k} p_w, \text{ is}$$

$$F_k = \begin{cases} 1 - \boldsymbol{\alpha}\mathbf{T}_1^k\mathbf{e} & ; \quad a_0 = 0 < k \leq a_1 \\ 1 - \boldsymbol{\alpha}\prod_{i=1}^{j-1}\mathbf{T}_i^{a_i - a_{i-1}}\mathbf{T}_j^{k - a_{j-1}}\mathbf{e} & ; \quad a_{j-1} < k \leq a_j \quad ; \quad 2 \leq j \leq n+1. \end{cases}$$

Proof.

- For $a_0 = 0 < k \leq a_1$

$$F_k = \sum_{w=1}^{k} p_w = \boldsymbol{\alpha}\sum_{w=1}^{k}\mathbf{T}_1^{w-1}\mathbf{T}_1^0 = \boldsymbol{\alpha}\left(\mathbf{I} - \mathbf{T}_1^k\right)\left(\mathbf{I} - \mathbf{T}_1\right)^{-1}\mathbf{T}_1^0 = \boldsymbol{\alpha}\left(\mathbf{I} - \mathbf{T}_1^k\right)\mathbf{e} = 1 - \boldsymbol{\alpha}\mathbf{T}_1^k\mathbf{e}$$

- For $a_{j-1} < k \leq a_j \quad ; \quad 2 \leq j \leq n+1$

$$F_k = \sum_{s=1}^{j} B_s, \text{ where}$$

$$B_1 = \sum_{w=1}^{a_1} p_w = \boldsymbol{\alpha}\sum_{w=1}^{a_1}\mathbf{T}_1^{w-1}\mathbf{T}_1^0 = 1 - \boldsymbol{\alpha}\mathbf{T}_1^{a_1}\mathbf{e}$$

and for $s = 2,\ldots, j-1$ ($j \geq 3$)

$$B_s = \sum_{w=a_{s-1}+1}^{a_s} p_w = \boldsymbol{\alpha} \prod_{i=1}^{s-1} \mathbf{T}_i^{a_i - a_{i-1}} \sum_{w=a_{s-1}+1}^{a_s} \mathbf{T}_s^{w-a_{s-1}-1} \mathbf{T}_s^0$$

$$= \boldsymbol{\alpha} \prod_{i=1}^{s-1} \mathbf{T}_i^{a_i - a_{i-1}} \left( \mathbf{I} - \mathbf{T}_s^{a_s - a_{s-1}} \right) \mathbf{e} = \boldsymbol{\alpha} \prod_{i=1}^{s-1} \mathbf{T}_i^{a_i - a_{i-1}} \mathbf{e} - \boldsymbol{\alpha} \prod_{i=1}^{s-1} \mathbf{T}_i^{a_i - a_{i-1}} \mathbf{T}_s^{a_s - a_{s-1}} \mathbf{e}$$

and

$$B_j = \sum_{w=a_{j-1}+1}^{k} p_w = \boldsymbol{\alpha} \prod_{i=1}^{j-1} \mathbf{T}_i^{a_i - a_{i-1}} \sum_{w=a_{j-1}+1}^{k} \mathbf{T}_j^{w-a_{j-1}-1} \mathbf{T}_j^0 =$$

$$= \boldsymbol{\alpha} \prod_{i=1}^{j-1} \mathbf{T}_i^{a_i - a_{i-1}} \left( \mathbf{I} - \mathbf{T}_j^{k - a_{j-1}} \right) \mathbf{e} = \boldsymbol{\alpha} \prod_{i=1}^{j-1} \mathbf{T}_i^{a_i - a_{i-1}} \mathbf{e} - \boldsymbol{\alpha} \prod_{i=1}^{j-1} \mathbf{T}_i^{a_i - a_{i-1}} \mathbf{T}_j^{k - a_{j-1}} \mathbf{e}$$

Therefore,

$$F_k = 1 - \boldsymbol{\alpha} \prod_{i=1}^{j-1} \mathbf{T}_i^{a_i - a_{i-1}} \mathbf{T}_j^{k - a_{j-1}} \mathbf{e}.$$ □

The reliability function, $s_k = P(X > k)$, is immediate from the previous result,

$$s_k = 1 - F_k = P(X > k) = \begin{cases} \boldsymbol{\alpha} \mathbf{T}_1^k \mathbf{e} & ; \quad a_0 = 0 < k \leq a_1 \\ \boldsymbol{\alpha} \prod_{i=1}^{j-1} \mathbf{T}_i^{a_i - a_{i-1}} \mathbf{T}_j^{k - a_{j-1}} \mathbf{e} & ; \quad a_{j-1} < k \leq a_j \quad ; \quad 2 \leq j \leq n+1. \end{cases}$$

*Probability-Generating Function: $M_X(z)$*

The probability-generating function is defined as $M_X(z) = E\left[z^X\right] = \sum_{k=1}^{\infty} p_k z^k$ for certain values $z$ where the expectation exists (the inverses of the matrices exist). This function for the *n* cut-points phase-type distribution is

$$M_X(z) = E\left[z^X\right]$$

$$= z\boldsymbol{\alpha} (\mathbf{I} - z\mathbf{T}_1)^{-1} \mathbf{T}_1^0 + \boldsymbol{\alpha} \sum_{i=1}^{n} \left[ \prod_{j=1}^{i} \mathbf{T}_j^{a_j - a_{j-1}} \right] z^{a_i + 1} \left( (\mathbf{I} - z\mathbf{T}_{i+1})^{-1} \mathbf{T}_{i+1}^0 - (\mathbf{I} - z\mathbf{T}_i)^{-1} \mathbf{T}_i^0 \right)$$

Proof.

We can express the probability-generating function as

$$M_X(z) = E\left[z^X\right] = \sum_{w=1}^{n+1} p_k z^k = \sum_{i=1}^{n+1} A_i$$

where,

$$A_1 = \sum_{w=1}^{a_1} p_w z^w = \boldsymbol{\alpha} \sum_{w=1}^{a_1} \mathbf{T}_1^{w-1} z^w \mathbf{T}_1^0 = z\boldsymbol{\alpha}\left(\mathbf{I} - z\mathbf{T}_1\right)^{-1} \mathbf{T}_1^0 - \boldsymbol{\alpha}\mathbf{T}_1^{a_1} z^{a_1+1} \left(\mathbf{I} - z\mathbf{T}_1\right)^{-1} \mathbf{T}_1^0,$$

for $i = 2, \ldots, n$

$$A_i = \sum_{w=a_{i-1}+1}^{a_i} p_w z^w = \boldsymbol{\alpha} \prod_{j=1}^{i-1} \mathbf{T}_j^{a_j - a_{j-1}} z^{a_{i-1}+1} \sum_{w=a_{i-1}+1}^{a_i} \left(z\, \mathbf{T}_i\right)^{w - a_{i-1} - 1} \mathbf{T}_i^0$$

$$= \boldsymbol{\alpha} \prod_{j=1}^{i-1} \mathbf{T}_j^{a_j - a_{j-1}} z^{a_{i-1}} \left(\mathbf{I} - z\mathbf{T}_i\right)^{-1} \mathbf{T}_i^0 - \boldsymbol{\alpha} \prod_{j=1}^{i} \mathbf{T}_j^{a_j - a_{j-1}} z^{a_i+1} \left(\mathbf{I} - z\mathbf{T}_j\right)^{-1} \mathbf{T}_i^0$$

and

$$A_{n+1} = \sum_{w=a_n+1}^{\infty} p_w z^w = \boldsymbol{\alpha} \prod_{j=1}^{n} \mathbf{T}_j^{a_j - a_{j-1}} z^{a_n+1} \sum_{w=a_n+1}^{\infty} \left(z\, \mathbf{T}_{n+1}\right)^{w - a_n - 1} \mathbf{T}_{n+1}^0 = \boldsymbol{\alpha} \prod_{j=1}^{n} \mathbf{T}_j^{a_j - a_{j-1}} z^{a_n+1} \left(\mathbf{I} - z\mathbf{T}_{n+1}\right)^{-1} \mathbf{T}_{n+1}^0.$$

We can observe that

$$\sum_{i=1}^{n} A_i = z\boldsymbol{\alpha}\left(\mathbf{I} - z\mathbf{T}_1\right)^{-1} \mathbf{T}_1^0 + \boldsymbol{\alpha} \sum_{i=1}^{n-1} \left[ \prod_{j=1}^{i} \mathbf{T}_j^{a_j - a_{j-1}} \right] z^{a_i+1} \left( \left(\mathbf{I} - z\mathbf{T}_{i+1}\right)^{-1} \mathbf{T}_{i+1}^0 - \left(\mathbf{I} - z\mathbf{T}_i\right)^{-1} \mathbf{T}_i^0 \right)$$

$$- \boldsymbol{\alpha} \prod_{j=1}^{n} \mathbf{T}_j^{a_j - a_{j-1}} z^{a_n+1} \left(\mathbf{I} - z\mathbf{T}_n\right)^{-1} \mathbf{T}_n^0$$

Finally,

$$M_X(z) = \sum_{i=1}^{n+1} A_i = \sum_{i=1}^{n} A_i + A_{n+1}$$

$$= z\boldsymbol{\alpha}\left(\mathbf{I} - z\mathbf{T}_1\right)^{-1} \mathbf{T}_1^0 + \boldsymbol{\alpha} \sum_{i=1}^{n} \left[ \prod_{j=1}^{i} \mathbf{T}_j^{a_j - a_{j-1}} \right] z^{a_i+1} \left( \left(\mathbf{I} - z\mathbf{T}_{i+1}\right)^{-1} \mathbf{T}_{i+1}^0 - \left(\mathbf{I} - z\mathbf{T}_i\right)^{-1} \mathbf{T}_i^0 \right)$$

□

*EXPECTATION: E[x]*

The expectation is worked out from the probability-generating function as $E[X] = \left. \dfrac{\partial M_X(z)}{\partial z} \right|_{z=1}$.

Then,

$$\frac{\partial M_X(z)}{\partial z} = \boldsymbol{\alpha}(\mathbf{I}-z\mathbf{T}_1)^{-1}\mathbf{T}_1^0 + z\boldsymbol{\alpha}\mathbf{T}_1(\mathbf{I}-z\mathbf{T}_1)^{-2}\mathbf{T}_1^0$$

$$+\boldsymbol{\alpha}\sum_{i=1}^{n}\left[\prod_{j=1}^{i}\mathbf{T}_j^{a_j-a_{j-1}}\right](a_i+1)z^{a_i}\left((\mathbf{I}-z\mathbf{T}_{i+1})^{-1}\mathbf{T}_{i+1}^0 - (\mathbf{I}-z\mathbf{T}_i)^{-1}\mathbf{T}_i^0\right)$$

$$+\boldsymbol{\alpha}\sum_{i=1}^{n}\left[\prod_{j=1}^{i}\mathbf{T}_j^{a_j-a_{j-1}}\right]z^{a_i+1}\left(\mathbf{T}_{i+1}(\mathbf{I}-z\mathbf{T}_{i+1})^{-2}\mathbf{T}_{i+1}^0 - \mathbf{T}_i(\mathbf{I}-z\mathbf{T}_i)^{-2}\mathbf{T}_i^0\right)$$

If it is evaluated in $z=1$, then

$$E[X] = \left.\frac{\partial M_X(z)}{\partial z}\right|_{z=1} = 1+\boldsymbol{\alpha}\mathbf{T}_1(\mathbf{I}-\mathbf{T}_1)^{-1}\mathbf{e}$$

$$+\boldsymbol{\alpha}\sum_{i=1}^{n}\left[\prod_{j=1}^{i}\mathbf{T}_j^{a_j-a_{j-1}}\right]\left(\mathbf{T}_{i+1}(\mathbf{I}-\mathbf{T}_{i+1})^{-1} - \mathbf{T}_i(\mathbf{I}-\mathbf{T}_i)^{-1}\right)\mathbf{e}$$

$$= \boldsymbol{\alpha}(\mathbf{I}-\mathbf{T}_1)^{-1}\mathbf{e}$$

$$+\boldsymbol{\alpha}\sum_{i=1}^{n}\left[\prod_{j=1}^{i}\mathbf{T}_j^{a_j-a_{j-1}}\right]\left((\mathbf{I}-\mathbf{T}_{i+1})^{-1} - (\mathbf{I}-\mathbf{T}_i)^{-1}\right)\mathbf{e},$$

given that

$$1+\boldsymbol{\alpha}\mathbf{T}_1(\mathbf{I}-\mathbf{T}_1)^{-1}\mathbf{e} = \boldsymbol{\alpha}(\mathbf{I}-\mathbf{T}_1)(\mathbf{I}-\mathbf{T}_1)^{-1}\mathbf{e} + \boldsymbol{\alpha}\mathbf{T}_1(\mathbf{I}-\mathbf{T}_1)^{-1}\mathbf{e}$$
$$= \boldsymbol{\alpha}(\mathbf{I}-\mathbf{T}_1)^{-1}\mathbf{e} - \boldsymbol{\alpha}\mathbf{T}_1(\mathbf{I}-\mathbf{T}_1)^{-1}\mathbf{e} + \boldsymbol{\alpha}\mathbf{T}_1(\mathbf{I}-\mathbf{T}_1)^{-1}\mathbf{e} = \boldsymbol{\alpha}(\mathbf{I}-\mathbf{T}_1)^{-1}\mathbf{e}$$

and

$$(\mathbf{I}-\mathbf{T}_{i+1})(\mathbf{I}-\mathbf{T}_{i+1})^{-1} = (\mathbf{I}-\mathbf{T}_{i+1})^{-1} - \mathbf{T}_{i+1}(\mathbf{I}-\mathbf{T}_{i+1})^{-1} = \mathbf{I},$$
$$\mathbf{T}_{i+1}(\mathbf{I}-\mathbf{T}_{i+1})^{-1} = (\mathbf{I}-\mathbf{T}_{i+1})^{-1} - \mathbf{I},$$

for any $i = 1,\ldots, n$. $\square$

In a similar way $E[X(X-1)] = \left.\frac{\partial^2 M_X(z)}{\partial z^2}\right|_{z=1}$ can be obtained,

$$E[X(X-1)] = 2\boldsymbol{\alpha}(I-\mathbf{T}_1)^{-2}\mathbf{T}_1\mathbf{e} + 2\boldsymbol{\alpha}\sum_{i=1}^{n}\left[\prod_{j=1}^{i}\mathbf{T}_j^{a_j-a_{j-1}}\right]\left[(\mathbf{I}-\mathbf{T}_{i+1})^{-1}\left(a_i\mathbf{I}+(\mathbf{I}-\mathbf{T}_{i+1})^{-1}\mathbf{T}_{i+1}\right)\right.$$
$$\left. -(\mathbf{I}-\mathbf{T}_i)^{-1}\left(a_i\mathbf{I}+2(\mathbf{I}-\mathbf{T}_i)^{-1}\right)\right]\mathbf{e}.$$

**Appendix C**

In this appendix the conditional expectations of $B_i$, $Z_{i,h}$, $N_{ij,h}$ and $N_{i,m+1,h}$ is worked out for $i, j = 1,\ldots,m$ and $h = 1,\ldots,n+1$. To simplify the notation we assume that the sample size is equal to one and consider a single Markov process $I(\cdot)$ corresponding to the phase-type distribution with absorption time $Y$. The conditioning expectations are equal to the following expressions.

*Conditioning expectation of $B_i$*

$$E_{(\boldsymbol{\alpha},\mathbf{T}_1,\ldots,\mathbf{T}_{n+1})}(B_i \mid Y = y) = \frac{P(I(0) = i, Y \in dy)}{P(Y \in dy)}$$

$$= \frac{P(I(0) = i) P(Y \in dy \mid I(0) = i)}{P(Y \in dy)}$$

$$= \frac{\alpha_i \mathbf{e}_i^T \mathbf{b}(y \mid \mathbf{T}_1,\ldots,\mathbf{T}_{n+1})}{\boldsymbol{\alpha} \mathbf{b}(y \mid \mathbf{T}_1,\ldots,\mathbf{T}_{n+1})} \quad , \quad i = 1,2,\ldots,m,$$

being $\mathbf{b}(y \mid \cdot)$ the backward row vector with the absorbing rates from the transient states at time $y$ and $\mathbf{e}_i$ a column vector of appropriate order whose elements are zero except the $i$-th which is equal to 1.

That is,

$$\mathbf{b}(y \mid \mathbf{T}_1,\ldots,\mathbf{T}_{n+1})$$

$$= \begin{cases} \exp(\mathbf{T}_1 y)\mathbf{T}_1^0 & ; \quad a_0 = 0 < y \leq a_1 \\ \exp(\mathbf{T}_1 a_1)\exp(\mathbf{T}_2(y - a_1))\mathbf{T}_2^0 & ; \quad a_1 < y \leq a_2 \\ \exp(\mathbf{T}_1 a_1)\prod_{i=2}^{j-1}\exp(\mathbf{T}_i(a_i - a_{i-1}))\exp(\mathbf{T}_j(y - a_{j-1}))\mathbf{T}_j^0 & ; \quad a_{j-1} < y \leq a_j \quad ; \quad 3 \leq j \leq n+1. \end{cases}$$

*Conditioning expectation of $Z_{i,h}$*

For $i = 1,\ldots, m$ and $h = 1,\ldots, n+1$,

$$E_{(\boldsymbol{\alpha},\mathbf{T}_1,...,\mathbf{T}_{n+1})}\left(Z_{i,h} \mid Y=y\right) = E\left[\int_{a_{h-1}}^{a_h} I_{\{I(t)=i\}} dt \mid Y=y\right]$$

$$= \int_{a_{h-1}}^{a_h} \frac{P\{I(t)=i, Y \in dy\}}{P\{Y \in dy\}} dt$$

$$= \frac{\int_{a_{h-1}}^{a_h} P\{I(t)=i\} P\{Y \in dy \mid I(t)=i\} dt}{P\{Y \in dy\}}$$

$$= \frac{c_{ii}(y, h \mid \boldsymbol{\alpha}, \mathbf{T}_1,...,\mathbf{T}_{n+1})}{\boldsymbol{\alpha} \mathbf{b}(y \mid \mathbf{T}_1,...,\mathbf{T}_{n+1})},$$

being $c_{ii}(y,h/\cdot)$ the mean time in state $i$ in the interval $(a_{h-1}, a_h]$ and failure at time $dy$. The general function $c_{ij}(y,h/\cdot)$ for any transient states $i, j = 1,\ldots, m$ is

$$c_{ij}(y, h \mid \boldsymbol{\alpha}, \mathbf{T}_1,...,\mathbf{T}_h) = \int_{a_{h-1}}^{\max(a_{h-1},\min(y,a_h))} \mathbf{a}(u \mid \boldsymbol{\alpha}, \mathbf{T}_1,...,\mathbf{T}_h) \mathbf{e}_i \mathbf{e}_j^T$$
$$\cdot \mathbf{f}\left(\max(a_{h-1}, \min(y, a_h)) - u, y, h \mid \mathbf{T}_1,...,\mathbf{T}_{n+1}\right) du$$

being

$$\mathbf{a}(y \mid \boldsymbol{\alpha}, \mathbf{T}_1,...,\mathbf{T}_{n+1})$$

$$= \begin{cases} \boldsymbol{\alpha} \exp(\mathbf{T}_1 y) & ; \; a_0 = 0 < y \le a_1 \\ \boldsymbol{\alpha} \exp(\mathbf{T}_1 a_1) \exp(\mathbf{T}_2 (y-a_1)) & ; \; a_1 < y \le a_2 \\ \boldsymbol{\alpha} \exp(\mathbf{T}_1 a_1) \prod_{i=2}^{j-1} \exp(\mathbf{T}_i (a_i - a_{i-1})) \exp(\mathbf{T}_j (y - a_{j-1})) & ; \; a_{j-1} < y \le a_j \; ; \; 3 \le j \le n+1 \end{cases}$$

and

$$\mathbf{f}(x, y, h \mid \mathbf{T}_1,...,\mathbf{T}_{n+1}) = \begin{cases} \exp(\mathbf{T}_h x) \mathbf{T}_h^0 & ; \; a_{h-1} < y \le a_h \\ \exp(\mathbf{T}_h x) \exp(\mathbf{T}_{h+1}(y - a_h)) \mathbf{T}_{h+1}^0 & ; \; a_h < y \le a_{h+1} \\ \exp(\mathbf{T}_h x) \prod_{i=h+1}^{j-1} \exp(\mathbf{T}_i (a_i - a_{i-1})) \exp(\mathbf{T}_j (y - a_{j-1})) \mathbf{T}_j^0 & ; \; a_{j-1} < y \le a_j \; ; \; h+1 < j \le n+1 \end{cases}$$

where $\mathbf{a}(u \mid \cdot)$ is the forward row vector with the probabilities of being in the corresponding transient state in the Markov process $I(\cdot)$ and $\mathbf{f}(x, y, h/\cdot)$ a column vector with the failure rates depending on the initial state.

*Conditioning expectation of $N_{ij,h}$*

$N_{ij,h}$ is the number of jumps between two non-absorbing states in the interval $(a_{h-1}, a_h]$. The conditional expectation can be worked out as follows for $i, j = 1, \ldots, m$ and $h = 1, \ldots, n+1$,

$$E_{(\boldsymbol{\alpha}, \mathbf{T}_1, \ldots, \mathbf{T}_{n+1})}\left[N_{ij,h} \big| Y = y\right] = \frac{\int_{a_{h-1}}^{a_h} P\{I(t) = i\} t_{ij,h} P\{Y \in dy | I(t) = j\} dt}{P\{Y \in dy\}}$$

$$= \frac{t_{ij,h} c_{ij}(y, h | \boldsymbol{\alpha}, \mathbf{T}_1, \ldots, \mathbf{T}_{n+1})}{\boldsymbol{\alpha} \mathbf{b}(y | \mathbf{T}_1, \ldots, \mathbf{T}_{n+1})}.$$

*Conditioning expectation of $N_{i,m+1,h}$*

$N_{i,m+1,h}$ is the number of jumps between the transient state $i$ to the absorbing state $m+1$ of the absorbing Markov process embedded $I(\cdot)$, in the interval $(a_{h-1}, a_h]$. The conditional expectation can be worked out as the conditional probability of jumping from the state $i$ to the absorbing state at a certain time in this interval. It follows for $i = 1, \ldots, m$ and $h = 1, \ldots, n+1$,

$$E_{(\boldsymbol{\alpha}, \mathbf{T}_1, \ldots, \mathbf{T}_{n+1})}\left(N_{i,m+1,h} | Y = y\right) = \lim_{\Delta \downarrow 0} \sum_{k=1}^{n+1} I_{\{a_{k-1} < y \leq a_k\}} \frac{P(Y \in dy, I(y - \Delta) = i)}{P(Y \in dy)}$$

$$= \lim_{\Delta \downarrow 0} \frac{\sum_{k=1}^{n+1} I_{\{a_{k-1} < y \leq a_k\}} P(I(y-\Delta) = i) P(Y \in dy | I(y-\Delta) = i)}{P(Y \in dy)}$$

$$= \lim_{\Delta \downarrow 0} \frac{\sum_{k=1}^{n+1} I_{\{a_{k-1} < y \leq a_k\}} \mathbf{a}(y - \Delta | \boldsymbol{\alpha}, \mathbf{T}_1, \ldots, \mathbf{T}_{n+1}) \mathbf{e}_i \mathbf{e}_i^T \exp(\mathbf{T}_h \Delta) \mathbf{T}_h^0}{P(Y \in dy)},$$

$$= \frac{\sum_{k=1}^{n+1} I_{\{a_{k-1} < y \leq a_k\}} t_{im+1,h} \mathbf{a}(y | \boldsymbol{\alpha}, \mathbf{T}_1, \ldots, \mathbf{T}_{n+1}) \mathbf{e}_i}{\boldsymbol{\alpha} \mathbf{b}(y | \mathbf{T}_1, \ldots, \mathbf{T}_{n+1})}$$

being $t_{im+1,h}$ the $i$-th element of the column vector $\mathbf{T}_h^0$.